\documentclass[seceq]{ptptex}
\usepackage{graphicx, wrapft, braket}
\input{colordvi.tex}

\title{Violation of the Rotational Invariance in the CMB Bispectrum}

\author{
Maresuke \textsc{Shiraishi}\footnote{Email: mare@a.phys.nagoya-u.ac.jp} 
and Shuichiro \textsc{Yokoyama}\footnote{Email: shu@a.phys.nagoya-u.ac.jp}
}
\inst{Department of Physics and Astrophysics, Nagoya University, Nagoya 464-8602, Japan}

\recdate{July 5, 2011; Revised September 21, 2011}

\abst{
We investigate a statistical anisotropy on the Cosmic Microwave Background (CMB) bispectrum, which can be generated from the primordial non-Gaussianity induced by quantum fluctuations of a vector field. 
We find new configurations in the multipole space of the CMB bispectrum
given by $\ell_1 = \ell_2 + \ell_3 + 2, |\ell_2 - \ell_3| - 2$ and their permutations, which violate the rotational invariance, such as an off-diagonal configuration in the CMB power spectrum. 
We also find that in a model presented by Yokoyama and Soda (2008), 
the amplitude of the statistically anisotropic bispectrum in the above configurations becomes as large as that in other configurations such as $\ell_1 = \ell_2 + \ell_3$.
As a result, it might be possible to detect these contributions in future experiments, which would give us novel information about the physics of the early Universe.}

\PTPindex{400, 435, 440, 442}  

\begin{document}
\maketitle

\section{Introduction}

The current cosmological observations, particularly 
Cosmic Microwave Background (CMB),
tell us that the Universe is almost isotropic, and 
primordial density fluctuations are almost Gaussian random fields.
However, in keeping with the progress of the experiments, 
 there have been many works that verify the possibility of the small deviation of the statistical isotropy, e.g., the so-called ``Axis of Evil''.
The analyses of the power spectrum by employing the current CMB data suggest that 
the deviation of the statistical isotropy is about $10 \%$ at most 
(e.g., Refs.~\citen{Groeneboom:2008fz, Groeneboom:2009cb, Frommert:2009qw,
Hanson:2009gu, Bennett:2010jb, Hanson:2010gu}).
Toward more precise measurements in future experiments,
there are a lot of theoretical discussions about
the effects of the statistical anisotropy on the CMB power spectrum,
 \cite{Ackerman:2007nb}\tocite{Gumrukcuoglu:2010yc} 
e.g., the presence of the off-diagonal configuration
of the multipoles in the CMB power spectrum, which vanishes in the isotropic spectrum.

As is well known, it might be difficult to explain such statistical anisotropy
in the standard inflationary scenario.
However, recently, 
there have been several works about the possibility of generating the
statistically anisotropic primordial density fluctuations
in order to introduce nontrivial dynamics of the vector field. 
\cite{Dimopoulos:2006ms}\tocite{Karciauskas:2011fp} 
In Ref.~\citen{Yokoyama:2008xw}, the authors considered 
a modified hybrid inflation model where a waterfall field couples not only with an inflaton field 
but also with a massless vector field.
They have shown that, owing to the effect of fluctuations of the vector field, the primordial density fluctuations
may have a small deviation from the statistical isotropy and also
the deviation from the Gaussian statistics.
If the primordial density fluctuations deviate from the Gaussian statistics,
they produces the non-zero higher order spectra (corresponding to higher order correlation functions), e.g., the bispectrum (3-point function), the trispectrum (4-point function) and so on.
Hence, 
in the model presented in Ref.~\citen{Yokoyama:2008xw},
we can expect that there are characteristic signals not only in the CMB power spectrum but also in the CMB bispectrum. 

With these motivations, in this work, we calculate the CMB
statistically anisotropic bispectrum sourced from the curvature
perturbations generated in the modified hybrid inflation scenario
proposed in Ref.~\citen{Yokoyama:2008xw}, on the basis of the useful formula presented in Ref.~\citen{Shiraishi:2010kd}.
Then, we find the peculiar configurations of the multipoles which never appear in the isotropic bispectrum, like off-diagonal components in the CMB power spectrum. 

This paper is organized as follows. In the next section, we briefly review the inflation model where the scalar waterfall field couples with the vector field and calculate the bispectrum of curvature perturbations based on Ref.~\citen{Yokoyama:2008xw}. In \S\ref{sec:CMB_bis}, we give an exact form of the CMB statistically anisotropic bispectrum and analyze its behavior by numerical computation. Finally, we devote the final section to the summary and discussion. 

Throughout this paper, we obey the definition of the Fourier transformation as 
\begin{eqnarray}
f(\mib{x}) \equiv \int \frac{d^3 \mib{k}}{(2 \pi)^3} \tilde{f}(\mib{k}) e^{i \mib{k} \cdot \mib{x}}~,
\end{eqnarray}
and a normalization as $M_{\rm pl} \equiv (8\pi G)^{-1/2} = 1$.

\section{Statistically anisotropic non-Gaussianity in curvature perturbations}

In this section, we briefly review the mechanism of generating the statistically anisotropic bispectrum induced by primordial curvature perturbations proposed in Ref.~\citen{Yokoyama:2008xw}, 
where the authors set the system like the hybrid inflation wherein there are two scalar fields:
inflaton $\phi$ and waterfall field $\chi$, and a vector field $A_\mu$ coupled with a waterfall field. 
The action is given by   
\begin{eqnarray}
S &=& \int dx^4 \sqrt{-g} 
\left[\frac{1}{2}R - \frac{1}{2}g^{\mu\nu}(\partial_{\mu} \phi
 \partial_\nu \phi + \partial_\mu \chi \partial_\nu \chi)
- V(\phi, \chi, A_\nu) \right. \nonumber \\
&&\left. \qquad\qquad\quad 
- \frac{1}{4} g^{\mu\nu} g^{\rho \sigma}f^2(\phi) F_{\mu \rho} F_{\nu \sigma}
\right]~. \label{eq:action} 
\end{eqnarray}
Here, $F_{\mu \nu} \equiv \partial_\mu A_\nu - \partial_\nu A_\mu$ is the field strength of the vector field
$A_\mu$, $V(\phi, \chi, A_\mu)$
is the potential of fields and 
$f(\phi)$ denotes a gauge coupling. 
To guarantee the isotropy of the background Universe,
we need the condition that the energy density of the vector field
is negligible in the total energy of the Universe and we assume a small expectation value of the vector field. 
Therefore, we neglect the effect of the vector field on the background dynamics and also the evolution of
the fluctuations of the inflaton. 
 In the standard hybrid inflation (only with the inflaton and the waterfall field),
the inflation suddenly ends owing to the tachyonic instability of the waterfall field, which
is triggered when the inflaton reaches a critical value $\phi_{\rm e}$. 
In the system described using Eq.~(\ref{eq:action}), 
however, $\phi_{\rm e}$ may fluctuate 
owing to the fluctuation of the vector field and it generates additional curvature perturbations. 

Using the $\delta N$ formalism \cite{Starobinsky:1982ee, Starobinsky:1986fxa, Sasaki:1995aw, Sasaki:1998ug, Lyth:2004gb, Lyth:2005fi}, 
the total curvature perturbation on the uniform-energy-density hypersurface at the end of inflation $t=t_{\rm e}$ can be estimated in terms of 
the perturbation of the $e$-folding number as  
\begin{eqnarray}
\zeta (t_{\rm e}) &=& \delta N (t_{\rm e}, t_*) \nonumber \\
&=& \frac{\partial N}{\partial \phi_*} \delta \phi_* 
+ \frac{1}{2} \frac{\partial^2 N}{\partial \phi_*^2} \delta \phi_*^2  
+ \frac{\partial N}{\partial \phi_{\rm e}} 
\frac{d \phi_{\rm e}(A)}{d A^\mu} \delta A_{\rm e}^\mu 
\nonumber \\
&& 
+ \frac{1}{2} 
\left[ \frac{\partial N}{\partial \phi_{\rm e}} 
\frac{d^2 \phi_{\rm e}(A)}{dA^\mu dA^\nu}
+ \frac{\partial^2 N}{\partial \phi_{\rm e}^2} 
\frac{d \phi_{\rm e}(A)}{d A^\mu} \frac{d \phi_{\rm e}(A)}{d A^\nu} \right] 
\delta A_{\rm e}^\mu \delta A_{\rm e}^\nu ~.
\end{eqnarray}
Here, $t_*$ is the time when the scale of interest crosses the
horizon during the slow-roll inflation. 
Assuming the sudden decay of all fields into radiations just after the inflation, the curvature perturbations on the uniform-energy-density hypersurface become constant after the inflation ends.
Hence, at the leading order, the power spectrum and the bispectrum of curvature perturbations are respectively derived as 
\begin{eqnarray}
\Braket{\prod_{n=1}^2 \zeta(\mib{k_n})} 
&=& (2\pi)^3 N_*^2 P_{\phi} (k_1) \delta\left(\sum_{n=1}^2 \mib{k_n}\right)
 \nonumber \\
&& + N_{\rm e}^2 \frac{d \phi_{\rm e}(A)}{d A^\mu} \frac{d \phi_{\rm e}(A)}{d A^\nu} 
\Braket{\delta A^\mu_{\rm e}(\mib{k_1}) \delta A^\nu_{\rm e}(\mib{k_2})} ~, \label{eq:ini_power_deltaN} \\
\Braket{\prod_{n=1}^3 \zeta(\mib{k_n})} 
&=& (2\pi)^3 N^2_* N_{**} [P_\phi(k_1) P_\phi(k_2) + 2 \ {\rm perms.}]
\delta\left(\sum_{n=1}^3 \mib{k_n}\right) \nonumber \\
&& + N_{\rm e}^3 \frac{d \phi_{\rm e}(A)}{d A^\mu} \frac{d \phi_{\rm e}(A)}{d A^\nu} \frac{d \phi_{\rm e}(A)}{d A^\rho} 
\Braket{\delta A^\mu_{\rm e}(\mib{k_1}) \delta A^\nu_{\rm
e}(\mib{k_2}) \delta A^\rho_{\rm e}(\mib{k_3})} \nonumber \\
&& + N_{\rm e}^4 \frac{d \phi_{\rm e}(A)}{d A^\mu} \frac{d \phi_{\rm e}(A)}{d A^\nu}
\left( \frac{1}{N_{\rm e}} \frac{d^2 \phi_{\rm e}(A)}{d A^\rho d A^\sigma} 
+ \frac{N_{\rm e e}}{N_{\rm e}^2} \frac{d \phi_{\rm e}(A)}{dA^\rho} 
\frac{d \phi_{\rm e}(A)}{dA^\sigma} \right) \nonumber \\
&& \times 
\left[ \Braket{\delta A^\mu_{\rm e}(\mib{k_1}) \delta A^\nu_{\rm e}(\mib{k_2})
(\delta A^\rho \star \delta A^\sigma)_{\rm e}(\mib{k_3})} + 2 \ {\rm perms.}\right]~, \label{eq:ini_bis_deltaN}
\end{eqnarray}
where $P_\phi(k) = H_*^2 / (2 k^3)$ is the power spectrum of the fluctuations of the inflaton, 
$N_{*} \equiv \partial N / \partial \phi_*$, 
$N_{**} \equiv \partial^2 N / \partial \phi_*^2$, 
$N_{\rm e} \equiv \partial N / \partial \phi_{\rm e}$, 
$N_{\rm ee} \equiv \partial^2 N / \partial \phi_{\rm e}^2
$, and 
$~\star~$ denotes the convolution.
Here, we assume that $\delta \phi_*$ is a Gaussian random field
and $\Braket{ \delta \phi A^\mu } = 0$.

For simplicity, we estimate the fluctuation of the vector fields in the Coulomb gauge: $\delta A_0 = 0$ and $k_i A^i = 0$. Then, the evolution equation of the fluctuations of the vector field is given by
\begin{eqnarray}
{\cal A}''_i - \frac{f''}{f} {\cal A}_i 
- a^2 \partial_j \partial^j {\cal A}_i = 0~,
\end{eqnarray}
where ${\cal A}_i \equiv f \delta A_i$, $~'~$
denotes the derivative with respect to the conformal time, and we neglect the contribution from the potential term. 
When $f \propto a, a^{-2}$ with appropriate
quantization of the fluctuations of the vector field, 
we have the scale-invariant power spectrum of $\delta A^i$ on superhorizon
scale as \cite{Yokoyama:2008xw, Dimopoulos:2009am, Martin:2007ue} 
\begin{eqnarray}
\Braket{\delta A^i_{\rm e}(\mib{k_1}) \delta A^j_{\rm e}(\mib{k_2})} 
= (2\pi)^3 P_\phi(k) f_{\rm e}^{-2} P^{ij}(\hat{\mib{k_1}})
\delta\left(\sum_{n=1}^2 \mib{k_n}\right)~, \label{eq:ini_power_A}
\end{eqnarray}
where $a$ is the scale factor, $P^{ij}(\hat{\mib{k}}) = \delta^{ij} -
\hat{k}^i \hat{k}^j$, $~\hat{}~$ denotes the unit vector, and $f_{\rm e}
\equiv f(t_{\rm e})$. 
Therefore, substituting this expression into Eq.~(\ref{eq:ini_power_deltaN}),
we can rewrite the power spectrum of the primordial curvature perturbations, $\zeta$, as
\begin{eqnarray}
\Braket{\prod_{n=1}^2 \zeta(\mib{k_n})} &\equiv& (2 \pi)^3 P_{\zeta}(\mib{k_1}) 
\delta\left(\sum_{n=1}^2 \mib{k_n} \right)~, \\
P_{\zeta}(\mib{k}) 
&=& P_\phi(k) \left[ N_*^2 + \left(\frac{N_{\rm e}}{f_{\rm e}}\right)^2 
q^i q^j P_{ij}(\hat{\mib{k}}) \right]~,
\end{eqnarray}
where $q_i \equiv d \phi_{\rm e} / d A^i, q_{ij} \equiv d^2 \phi_{\rm e} / (d A^i d A^j)$.
We can divide this expression into the isotropic part and
the anisotropic part as~\cite{Ackerman:2007nb} 
\begin{eqnarray}
P_\zeta(\mib{k}) \equiv P_\zeta^{\rm iso}(k)
\left[ 1 + g_\beta \left( \hat{\mib{q}} \cdot \hat{\mib{k}} \right)^2
\right]~, \label{eq:P} 
\end{eqnarray}
with
\begin{eqnarray}
P^{\rm iso}_\zeta(k) = N_*^2 P_\phi(k) (1 + \beta)~, \ \ 
g_\beta = - \frac{\beta}{1 + \beta}~,
\end{eqnarray}
where $\beta = \left(N_{\rm e} / N_* / f_{\rm e}\right)^2 |\mib{q}|^2$.
The
bispectrum of the primordial curvature perturbation given by Eq.~(\ref{eq:ini_bis_deltaN}) 
can be written as
\begin{eqnarray}
\Braket{\prod_{n=1}^3 \zeta(\mib{k_n})} &\equiv& (2 \pi)^3 F_{\zeta}(\mib{k_1}, \mib{k_2}, \mib{k_3}) 
\delta\left(\sum_{n=1}^3 \mib{k_n} \right)~, \label{eq:ini_bis_general} \\
F_\zeta(\mib{k_1}, \mib{k_2}, \mib{k_3}) 
&=& \left( \frac{g_\beta}{\beta} \right)^2 
P^{\rm iso}_\zeta(k_1) P^{\rm iso}_\zeta(k_2) \nonumber \\
&&\times \left[\frac{N_{**}}{N_*^2} + \beta^2 \hat{q}^a \hat{q}^b
\left(\frac{1}{N_{\rm e}} \hat{q}^{cd}  + \frac{N_{\rm ee}}{N_{\rm e}^2}\hat{q}^c \hat{q}^d \right)
P_{ac}(\hat{\mib{k_1}}) P_{bd}(\hat{\mib{k_2}}) \right] 
\nonumber \\
&&
+ 2 \ {\rm perms.}~. \label{eq:F_general}
\end{eqnarray}
Here, $\hat{q}^{cd} \equiv q^{cd} / |\mib{q}|^2$ and we have assumed that
the fluctuation of the vector field $\delta A^i$ almost obeys Gaussian statistics; hence, $\Braket{ \delta A_{\rm e}^\mu(\mib{k_1}) \delta A_{\rm e}^\nu(\mib{k_2}) \delta A_{\rm e}^\rho(\mib{k_3}) } = 0$. 

Hereinafter, for calculating the CMB bispectrum explicitly, we adopt
a simple model whose potential looks like an Abelian Higgs model in the unitary gauge as~\cite{Yokoyama:2008xw}
\begin{eqnarray}
V(\phi, \chi, A^i) = \frac{\lambda}{4}(\chi^2 - v^2)^2 + \frac{1}{2} g^2
 \phi^2 \chi^2 + \frac{1}{2}m^2 \phi^2 + \frac{1}{2}h^2 A^\mu A_\mu \chi^2~,
\end{eqnarray}
where $\lambda, g$, and $h$ are the coupling constants, $m$
is the inflaton mass, and $v$ is the vacuum expectation value of $\chi$. 
Since the effective mass squared of the waterfall field is given by 
\begin{eqnarray} 
m_\chi^2 
\equiv \frac{\partial^2 V}{\partial \chi^2} 
= - \lambda v^2 + g^2 \phi_{\rm e}^2 + h^2 A^i A_i = 0~,
\end{eqnarray}
and the critical value of the inflaton $\phi_{\rm e}$ can be obtained as
\begin{eqnarray}
g^2 \phi_{\rm e}^2 = \lambda v^2 - h^2 A^iA_i~,
\end{eqnarray}
we can express $\beta$, $q^i$, and $q^{ij}$ in Eq.~(\ref{eq:F_general})
in terms of the model parameters as 
\begin{eqnarray}
\hat{q}^i = - \hat{A}^i~, \ \ \ 
\hat{q}^{ij} = - \frac{1}{\phi_{\rm e}}
\left[\left( \frac{g \phi_{\rm e}}{h A} \right)^2 \delta^{ij} + \hat{A}^i
 \hat{A}^j \right]~, \ \ \ 
\beta \simeq \frac{1}{f_{\rm e}^2} \left( \frac{h^2 A}{g^2 \phi_{\rm e}} \right)^2~,  
\end{eqnarray}
where we have used $N_* \simeq - N_e \simeq 1/ \sqrt{2\epsilon}$
with 
$\epsilon \equiv (\partial V / \partial \phi / V)^2 / 2$ being a slow-roll parameter and $|\mib{A}| \equiv A$.
Substituting these quantities into Eq.~(\ref{eq:F_general}), the bispectrum of primordial curvature
perturbations is obtained as
\begin{eqnarray}
F_\zeta (\mib{k_1}, \mib{k_2}, \mib{k_3}) 
&=& C P^{\rm iso}_\zeta(k_1) P^{\rm iso}_\zeta(k_2) 
\hat{A}^a \hat{A}^b
\delta^{cd}
P_{ac}(\hat{\mib{k_1}}) P_{bd}(\hat{\mib{k_2}})
+ 2 \ {\rm perms.}~, \label{eq:F} \\
C &\equiv& - g^2_\beta \frac{\phi_{\rm e}}{N_{\rm e}} \left( \frac{g}{h A} \right)^2~.
\end{eqnarray}
Note that in the above expression, we have neglected the effect of the longitudinal polarization in the vector field for simplicity
\footnote{Owing to this treatment, we can use the quantities estimated in the Coulomb gauge as Eq.~(\ref{eq:F_general}). In a more precise discussion, we should take into account the contribution of the longitudinal mode in the unitary gauge.} and the terms that are
suppressed by a slow-roll parameter $\eta \equiv \partial^2 V /
\partial \phi ^2 / V $ because $- N_{**} / N_*^2 \simeq N_{\rm ee} /
N_{\rm e}^2 \simeq - (N_{\rm e} \phi_{\rm e})^{-1} \simeq \eta$.
Since the current CMB observations suggest $g_\beta < {\cal O}(0.1)$
(e.g., Refs.~\citen{Groeneboom:2008fz, Groeneboom:2009cb}) and $N_{\rm
e}^{-1} \simeq - \sqrt{2 \epsilon}$, the overall amplitude of the bispectrum in this model, $C$, does not seem to be sufficiently large to
be detected.    
However, even if $g_\beta \ll 1$ and $\epsilon \ll 1$, $C$ can become greater than unity for $ (g / h A)^2 \phi_{\rm e} \gg 1$. Thus, we expect meaningful signals also in the CMB bispectrum. 
Then, in the next section,
we closely investigate the CMB bispectrum generated from the primordial bispectrum given by Eq.~(\ref{eq:F})
and discuss a new characteristic feature of the CMB bispectrum induced by the statistical anisotropy of the primordial bispectrum. 

\section{CMB statistically anisotropic bispectrum}\label{sec:CMB_bis}

In this section, we give a formula of the CMB bispectrum generated from the
primordial bispectrum, which has statistical anisotropy owing to the fluctuations of the vector field,
given by Eq.~(\ref{eq:F}).
We also discuss the special signals of this CMB bispectrum, which vanish in the statistically isotropic bispectrum. 

\subsection{Formulation}

The CMB fluctuation can be expanded in terms of the spherical harmonic
function as
\begin{eqnarray}
\frac{\Delta X}{X} = \sum_{\ell m} a_{X, \ell m} Y_{\ell m}(\hat{\mib{n}})~,
\end{eqnarray} 
where $\hat{\mib{n}}$ is a unit vector pointing toward a line-of-sight direction, and $X$ denotes the intensity ($\equiv I$) and polarizations ($\equiv E, B$). 
The coefficient, $a_{\ell m}$, generated from primordial curvature
perturbations, $\zeta$, is expressed as \cite{Shiraishi:2010kd, Shiraishi:2010sm}
\begin{eqnarray}
a_{X, \ell m} &=& 4\pi (-i)^\ell \int_0^\infty \frac{k^2
 dk}{(2\pi)^3}
\zeta_{\ell m}(k) {\cal T}_{X, \ell}(k)~, \ \ \ ({\rm for} \  X = I,E)  \\ 
\zeta_{\ell m}(k)  &\equiv& \int d^2 \hat{\mib{k}} \zeta(\mib{k})
Y^*_{\ell m}(\hat{\mib{k}})~, \label{eq:zeta_lm}
\end{eqnarray}
where ${\cal T}_{X, \ell}$ is the time-integrated transfer function of scalar
modes as calculated in Refs.~\citen{Hu:1997hp} and \citen{Zaldarriaga:1996xe}. Using
these equations, the CMB bispectrum generated from the bispectrum of the primordial
curvature perturbations is given by
\begin{eqnarray}
\Braket{\prod_{n=1}^3 a_{X_n, \ell_n m_n}}
= \left[\prod_{n=1}^3 4 \pi (-i)^{\ell_n} \int_0^\infty \frac{k_n^2
 dk_n}{(2\pi)^3} {\cal T}_{X_n, \ell_n}(k_n) \right]
\Braket{\prod_{n=1}^3 \zeta_{\ell_n m_n}(k_n)} , \label{eq:form_CMB_bis}
\end{eqnarray}
with
\begin{eqnarray}
\Braket{\prod_{n=1}^3 \zeta_{\ell_n m_n}(k_n)} 
&=& \left[ \prod_{n=1}^3 \int d^2 \hat{\mib{k_n}} Y^*_{\ell_n
     m_n}(\hat{\mib{k_n}}) \right] \nonumber \\
&&\times 
(2\pi)^3 \delta\left(\sum_{n=1}^3 \mib{k_n} \right) 
F_\zeta(\mib{k_1}, \mib{k_2}, \mib{k_3})~. 
\end{eqnarray}
We expand the angular dependences that appear in the Dirac delta function, $\delta (\mib{k_1}+\mib{k_2}+\mib{k_3})$, and
the function, $F_\zeta(\mib{k_1},\mib{k_2},\mib{k_3})$, given by Eq.~(\ref{eq:F}) with respect to the spin spherical harmonics as 
\begin{eqnarray}
\delta\left( \sum_{n=1}^3 \mib{k_n} \right) 
&=& 8 \int_0^\infty y^2 dy 
\left[ \prod_{n=1}^3 \sum_{L_n M_n} 
 (-1)^{L_n/2} j_{L_n}(k_n y) 
Y_{L_n M_n}^*(\hat{\mib{k_n}}) \right] 
\nonumber \\
&&\times 
I_{L_1 L_2 L_3}^{0 \ 0 \ 0}
 \left(
  \begin{array}{ccc}
  L_1 & L_2 & L_3 \\
  M_1 & M_2 & M_3 
  \end{array}
 \right)~, \\
\hat{A}^a \hat{A}^b \delta^{cd} 
P_{ac}(\hat{\mib{k_1}}) P_{bd}(\hat{\mib{k_2}}) 
&=& - 4\left(\frac{4\pi}{3}\right)^3 
\sum_{L, L',L_A = 0,2} I_{L 1 1}^{0 1 -1} I_{L' 1 1}^{0 1
-1} 
I_{1 1 L_A}^{0 0 0} 
\left\{
  \begin{array}{ccc}
  L & L' & L_A \\
  1 & 1 & 1
  \end{array}
 \right\} \nonumber \\
&&\times \sum_{M M' M_A} Y_{L M}^*(\hat{\mib{k_1}}) Y_{L' M'}^*(\hat{\mib{k_2}}) 
Y_{L_A M_A}^*(\hat{\mib{A}})  
\nonumber \\
&&\times 
\left(
  \begin{array}{ccc}
  L & L' & L_A \\
  M & M' & M_A 
  \end{array}
 \right) ~, \label{eq:product}
\end{eqnarray}
where the $2 \times 3$ matrices of a bracket and a curly bracket denote the Wigner-$3j$ and $6j$
symbols, respectively, and 
\begin{eqnarray}
I^{s_1 s_2 s_3}_{l_1 l_2 l_3} 
\equiv \sqrt{\frac{(2 l_1 + 1)(2 l_2 + 1)(2 l_3 + 1)}{4 \pi}}
\left(
  \begin{array}{ccc}
  l_1 & l_2 & l_3 \\
   s_1 & s_2 & s_3 
  \end{array}
 \right)~.
\end{eqnarray}
Here, we have used the expressions of an arbitrary unit vector and a projection tensor as
\begin{eqnarray}
\hat{r}_a &=& 
\left(
  \begin{array}{c}
  \sin \theta_r \cos \phi_r \\
  \sin \theta_r \sin \phi_r \\
  \cos \theta_r
  \end{array}
 \right)
= \sum_m \alpha_a^m Y_{1 m}(\hat{\mib{r}})~, \\ 
P_{ab}(\hat{\mib{r}}) &=& \delta_{ab} - \hat{r}_a \hat{r}_b \nonumber \\
&=& -2 \sum_{L=0,2} I_{L 1 1}^{0 1 -1} \sum_{M m_a m_b} 
Y^*_{L M}(\hat{\mib{r}}) \alpha_a^{m_a} \alpha_b^{m_b} 
\left(
  \begin{array}{ccc}
  L & 1 & 1 \\
  M & m_a & m_b
  \end{array}
 \right) ~, \label{eq:projection_tensor}
\end{eqnarray}
with 
\begin{eqnarray}
 \alpha_a^m &\equiv& \sqrt{\frac{2 \pi}{3}}
 \left(
  \begin{array}{ccc}
   -m (\delta_{m,1} + \delta_{m,-1}) \\
   i~ (\delta_{m,1} + \delta_{m,-1}) \\
   \sqrt{2} \delta_{m,0}
  \end{array}
\right)~, \label{eq:alpha}
\end{eqnarray}
and summation rules of the Wigner symbols as discussed in the Appendix of Ref.~\citen{Shiraishi:2010kd}.
\footnote{Equation (\ref{eq:projection_tensor}) is easily derived by
using the expression with a divergenceless vector described in
Ref.~\citen{Shiraishi:2011fi}. 
Equation (\ref{eq:alpha}) leads to the orthogonality relation as 
\begin{eqnarray}
\alpha^a_m \alpha_a^{m'} = \frac{4 \pi}{3} (-1)^m \delta_{m,-m'}~.
\end{eqnarray}
}
Note that for $Y_{00}^*(\hat{\mib{A}}) = 1/\sqrt{4\pi}$, the contribution
of $L_A = 0$ in Eq.~(\ref{eq:product}) is independent of the direction
of the vector field. Therefore, the statistical
anisotropy is generated from the signals of $L_A = 2$. 
By integrating these spherical harmonics over each unit vector, the angular
dependences on $\mib{k_1}, \mib{k_2}, \mib{k_3}$ can be reduced to the Wigner-$3j$ symbols as
\begin{eqnarray}
\int d^2 \hat{\mib{k_1}} Y^*_{\ell_1 m_1} Y^*_{L_1 M_1} Y^*_{L M}
&=& I_{\ell_1 L_1 L}^{0 \ 0 \ 0 }
 \left(
  \begin{array}{ccc}
  \ell_1 & L_1 & L \\
  m_1 & M_1 & M 
  \end{array}
 \right)~, \\
\int d^2 \hat{\mib{k_2}} Y^*_{\ell_2 m_2} Y^*_{L_2 M_2} Y^*_{L' M'}
&=& I_{\ell_2 L_2 L'}^{0 \ 0 \ 0 }
 \left(
  \begin{array}{ccc}
  \ell_2 & L_2 & L' \\
  m_2 & M_2 & M' 
  \end{array}
 \right)~,  \\
\int d^2 \hat{\mib{k_3}} Y^*_{\ell_3 m_3} Y^*_{L_3 M_3}
&=& (-1)^{m_3} \delta_{L_3, \ell_3} \delta_{M_3, -m_3} ~.
\end{eqnarray}
From these equations, we obtain an alternative explicit form of the
bispectrum of $\zeta_{\ell m}$ as 
\begin{eqnarray}
\Braket{\prod_{n=1}^3 \zeta_{\ell_n m_n}(k_n)}
&=& - (2\pi)^3 8 \int_0^\infty y^2 dy \sum_{L_1 L_2}
(-1)^{\frac{L_1+L_2+\ell_3}{2}} I_{L_1 L_2 \ell_3}^{0 \ 0 \ 0} \nonumber \\
&&\times P^{\rm iso}_\zeta(k_1) j_{L_1}(k_1 y) P^{\rm iso}_\zeta(k_2) j_{L_2}(k_2 y) C j_{\ell_3}(k_3
y) \nonumber \\
&&\times 
 4\left(\frac{4\pi}{3}\right)^3 (-1)^{m_3}
\sum_{L, L',L_A = 0,2} 
I_{L 1 1}^{0 1 -1} I_{L' 1 1}^{0 1 -1} \nonumber \\
&&\times I_{\ell_1 L_1 L}^{0 \ 0 \ 0 } I_{\ell_2 L_2 L'}^{0 \ 0 \ 0 }
I_{1 1 L_A}^{0 0 0} 
\left\{
  \begin{array}{ccc}
  L & L' & L_A \\
  1 & 1 & 1
  \end{array}
 \right\}
\nonumber \\
&&\times
 \sum_{M_1 M_2 M M' M_A} Y_{L_A M_A}^*(\hat{\mib{A}}) 
\left(
  \begin{array}{ccc}
  L_1 & L_2 & \ell_3 \\
  M_1 & M_2 & -m_3 
  \end{array}
 \right) \nonumber \\
&&\times
 \left(
  \begin{array}{ccc}
  \ell_1 & L_1 & L \\
  m_1 & M_1 & M 
  \end{array}
 \right)
 \left(
  \begin{array}{ccc}
  \ell_2 & L_2 & L' \\
  m_2 & M_2 & M' 
  \end{array}
 \right)
\left(
  \begin{array}{ccc}
  L & L' & L_A \\
  M & M' & M_A 
  \end{array}
 \right) \nonumber \\
&&+ {\rm 2 \ perms.} ~. \label{eq:zeta_lm_bis}
\end{eqnarray}
This equation implies that, owing to the vector field $\mib{A}$, 
the CMB bispectrum has a direction dependence, and hence, the dependence on $m_1, m_2, m_3$ cannot be confined 
only to a Wigner-$3j$ symbol, namely, 
\begin{eqnarray}
\Braket{\prod_{n=1}^3 \zeta_{\ell_n m_n}(k_n)} 
\neq (2\pi)^3 {\cal F}_{\ell_1 \ell_2 \ell_3}(k_1, k_2, k_3)
\left(
  \begin{array}{ccc}
  \ell_1 & \ell_2 & \ell_3 \\
  m_1 & m_2 & m_3 
  \end{array}
 \right)~. 
\end{eqnarray}
This fact truly indicates the violation of the rotational invariance in the 
bispectrum of the primordial curvature perturbations and leads to the
statistical anisotropy on the CMB bispectrum.

Let us consider the explicit form of the CMB bispectrum.
Here, we set the coordinate as $\hat{\mib{A}} = \hat{\mib{z}}$. Then, by substituting Eq.~(\ref{eq:zeta_lm_bis}) into Eq.~(\ref{eq:form_CMB_bis}) and using the
relation $Y_{L_A M_A}^*(\hat{\mib{z}}) = \sqrt{(2L_A + 1 )/ (4\pi)}
\delta_{M_A,0}$, the CMB bispectrum is expressed as 
\begin{eqnarray}
\Braket{\prod_{n=1}^3 a_{X_n, \ell_n m_n}} 
&=& - \int_0^\infty y^2 dy 
\left[ \prod_{n=1}^3 \frac{2}{\pi} \int_0^\infty k_n^2 dk_n 
{\cal T}_{X_n,\ell_n}(k_n) \right] \nonumber \\
&&\times \sum_{L_1 L_2}
(-1)^{\frac{\ell_1+\ell_2+L_1+L_2}{2} + \ell_3}
I_{L_1 L_2 \ell_3}^{0 \ 0 \ 0} 
\nonumber \\
&&\times 
P^{\rm iso}_\zeta(k_1) j_{L_1}(k_1 y) P^{\rm iso}_\zeta(k_2) j_{L_2}(k_2 y) C j_{\ell_3}(k_3
y)  \nonumber \\
&&\times 
4\left(\frac{4\pi}{3}\right)^3 (-1)^{m_3}
\sum_{L, L',L_A = 0,2}  
I_{L 1 1}^{0 1 -1} I_{L' 1 1}^{0 1 -1} \nonumber \\
&&\times I_{\ell_1 L_1 L}^{0 \ 0 \ 0 } I_{\ell_2 L_2 L'}^{0 \ 0 \ 0 }
I_{1 1 L_A}^{0 0 0} 
\left\{
  \begin{array}{ccc}
  L & L' & L_A \\
  1 & 1 & 1
  \end{array}
 \right\}
\nonumber \\
&&\times
\sqrt{\frac{2L_A +1}{4\pi}}
\sum_{M = -2}^2   
\left(
  \begin{array}{ccc}
  L_1 & L_2 & \ell_3 \\
  -m_1-M & -m_2+M & -m_3 
  \end{array}
 \right) \nonumber \\
&&\times 
 \left(
  \begin{array}{ccc}
  \ell_1 & L_1 & L \\
  m_1 & -m_1-M & M 
  \end{array}
 \right) 
 \left(
  \begin{array}{ccc}
  \ell_2 & L_2 & L' \\
  m_2 & -m_2+M & -M 
  \end{array}
 \right) \nonumber \\
&&\times 
\left(
  \begin{array}{ccc}
  L & L' & L_A \\
  M & -M & 0
  \end{array}
 \right) + {\rm 2 \ perms.}~. \label{eq:cmb_bis}
\end{eqnarray}
By taking into account the selection rules of the Wigner symbols
\cite{Shiraishi:2010kd}, the multipoles and azimuthal quantum
numbers are limited as
\begin{eqnarray}
&& \sum_{n=1}^3 \ell_n = {\rm even} ~, \ \
\sum_{n=1}^3 m_n = 0 ~, \nonumber \\
&& L_1 = |\ell_1 -2|, \ell_1, \ell_1 + 2~, \ \ 
L_2 = |\ell_2 -2|, \ell_2, \ell_2 + 2~, \nonumber \\
&& |L_2 - \ell_3| \leq L_1 \leq L_2 + \ell_3~, \label{eq:lm_range}
\end{eqnarray}
and the two permutations of $\ell_1, \ell_2, \ell_3$.

\subsection{Behavior of the CMB statistically anisotropic bispectrum}

On the basis of Eq.~(\ref{eq:cmb_bis}), we compute the CMB bispectra for the several $\ell$'s and $m$'s. 
Then, we modify the Boltzmann Code for Anisotropies in the Microwave Background (CAMB)
\cite{Lewis:1999bs,Lewis:2004ef} and use the Common Mathematical Library SLATEC \cite{slatec}. 

\begin{figure}[t]
  \begin{tabular}{cc}
    \begin{minipage}{0.5\hsize}
  \begin{center}
    \includegraphics[width=7.3cm,height=5.5cm,clip]{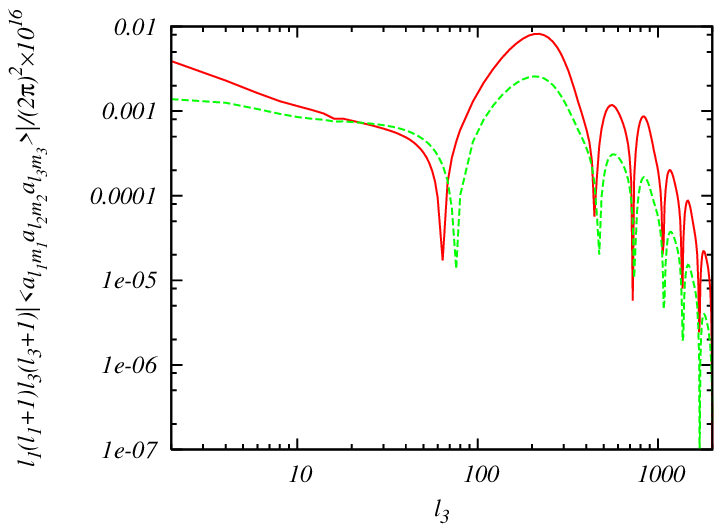}
  \end{center}
\end{minipage}
\begin{minipage}{0.5\hsize}
  \begin{center}
    \includegraphics[width=7.3cm,height=5.5cm,clip]{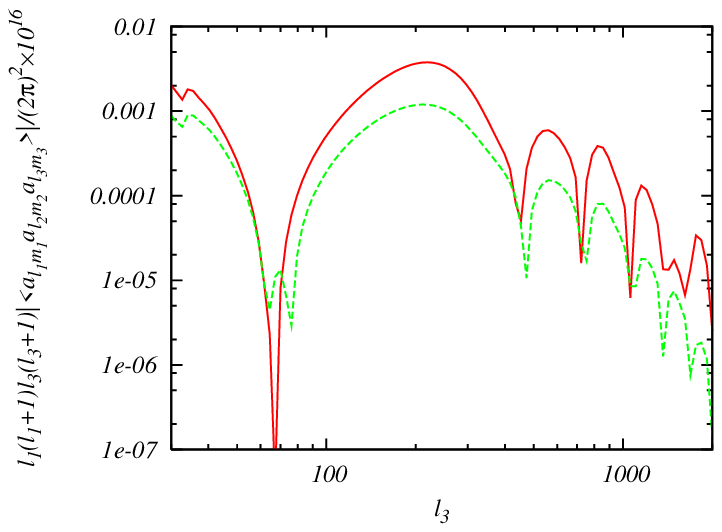}
  \end{center}
\end{minipage}
\end{tabular}
  \caption{(color online) Absolute values of the CMB statistically anisotropic
 bispectrum of the intensity mode given by Eq.~(\ref{eq:cmb_bis}) with $C = 1$ (red solid line) and the statistically isotropic one given by Eq.~(\ref{eq:cmb_bis_iso}) with $f_{\rm NL} = 5$ (green dashed line) for $\ell_1 = \ell_2 = \ell_3$. The left and right figures are plotted in the configurations $(m_1, m_2, m_3) = (0,0,0), (10,20,-30)$, respectively. 
The parameters are fixed to the mean values limited from the WMAP-7yr data as reported in Ref.~\citen{Komatsu:2010fb}.} \label{fig:SSS_RV_samel.eps}
\end{figure}

In Fig.~\ref{fig:SSS_RV_samel.eps}, 
the red solid lines are the CMB statistically anisotropic bispectra of the intensity mode given by
Eq.~(\ref{eq:cmb_bis}) with $C = 1$, and the green dashed lines are 
the statistically isotropic one sourced from the local-type non-Gaussianity of curvature perturbations given by \cite{Komatsu:2001rj}
\begin{eqnarray}
\Braket{\prod_{n=1}^3 a_{X_n, \ell_n m_n}} 
&=& I_{\ell_1 \ell_2 \ell_3}^{0~0~0}
\left(
  \begin{array}{ccc}
  \ell_1 & \ell_2 & \ell_3 \\
  m_1 & m_2 & m_3
  \end{array}
 \right) \nonumber \\
&&\times \int_0^\infty y^2 dy 
\left[ \prod_{n=1}^3 \frac{2}{\pi} \int_0^\infty k_n^2 dk_n 
{\cal T}_{X_n,\ell_n}(k_n) j_{\ell_n}(k_n y) \right] 
\nonumber \\
&&\times 
\left( P^{\rm iso}_\zeta(k_1) P^{\rm iso}_\zeta(k_2) \frac{6}{5} f_{\rm NL}
 + {\rm 2 \ perms.} \right)~, \label{eq:cmb_bis_iso}
\end{eqnarray} 
with $f_{\rm NL} = 5$ for $\ell_1 = \ell_2 = \ell_3$ and two sets of $m_1, m_2, m_3$. 
From this figure, we can see that the red solid lines are in good agreement with the green dashed
line in the dependence on $\ell$ for both configurations of $m_1, m_2, m_3$.
This seems to be because the bispectrum of primordial curvature perturbations affected by the fluctuations of vector field given by Eq.~(\ref{eq:F})
has not only the anisotropic part but also the isotropic part and both parts have the same amplitude. 
In this sense, it is expected that the angular dependence on the vector field $\hat{\mib{A}}$ does not contribute much to a change in the shape of the CMB bispectrum. 
We also find that the anisotropic bispectrum for $C \sim 0.3$ 
is comparable in magnitude to the case with $f_{\rm NL} = 5$ for the standard local type,
which corresponds to the upper bound on the local-type non-Gaussianity expected from the PLANCK experiment \cite{:2006uk}.

In the discussion of the CMB power spectrum,
if the rotational invariance is violated in the primordial power spectrum given by Eq.~(\ref{eq:P}),
the signals in the off-diagonal configurations of $\ell$ also have
nonzero values \cite{Ackerman:2007nb, Boehmer:2007ut, Watanabe:2010bu}.
Likewise,
there are special configurations in the CMB bispectrum induced from the
statistical anisotropy on the primordial bispectrum as Eq.~(\ref{eq:F}). 
The selection rule (\ref{eq:lm_range}) suggests
that the statistically anisotropic bispectrum (\ref{eq:cmb_bis}) could be nonzero in the multipole configurations given by 
\begin{eqnarray}
\ell_1 = |\ell_2 - \ell_3| - 4, |\ell_2 - \ell_3| - 2, \ell_2 + \ell_3 +
 2, \ell_2 + \ell_3 + 4~, \label{eq:special_l}
\end{eqnarray}
and two permutations of $\ell_1, \ell_2, \ell_3$. 
In contrast, in these configurations, 
the isotropic bispectrum (e.g., Eq.~(\ref{eq:cmb_bis_iso})) vanishes owing to the triangle condition of the
Wigner-$3j$ symbol 
$\left(
  \begin{array}{ccc}
  \ell_1 & \ell_2 & \ell_3 \\
  m_1 & m_2 & m_3
  \end{array}
 \right) $ and the nonzero components arise only from 
\begin{eqnarray}
|\ell_2 - \ell_3| \leq \ell_1 \leq \ell_2 + \ell_3~. \label{eq:default_l}
\end{eqnarray}
Therefore, the signals of the configurations (\ref{eq:special_l}) have the pure information of the statistical anisotropy on the CMB bispectrum. 

\begin{figure}[t]
  \begin{tabular}{cc}
    \begin{minipage}{0.5\hsize}
  \begin{center}
    \includegraphics[width=7.3cm,height=5.5cm,clip]{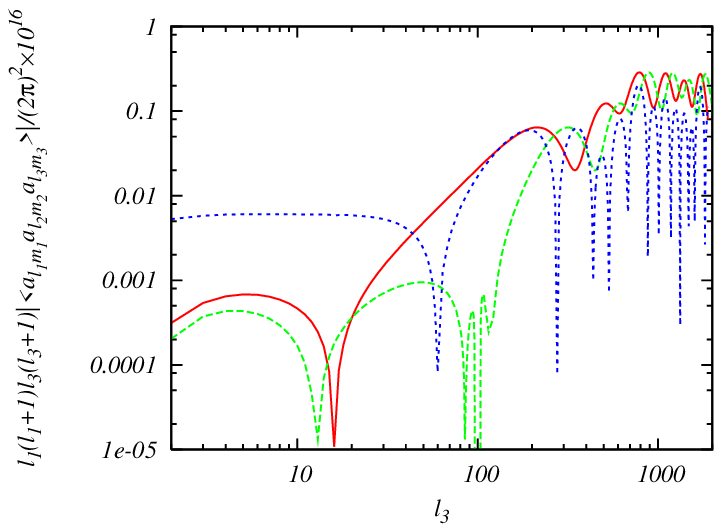}
  \end{center}
\end{minipage}
\begin{minipage}{0.5\hsize}
  \begin{center}
    \includegraphics[width=7.3cm,height=5.5cm,clip]{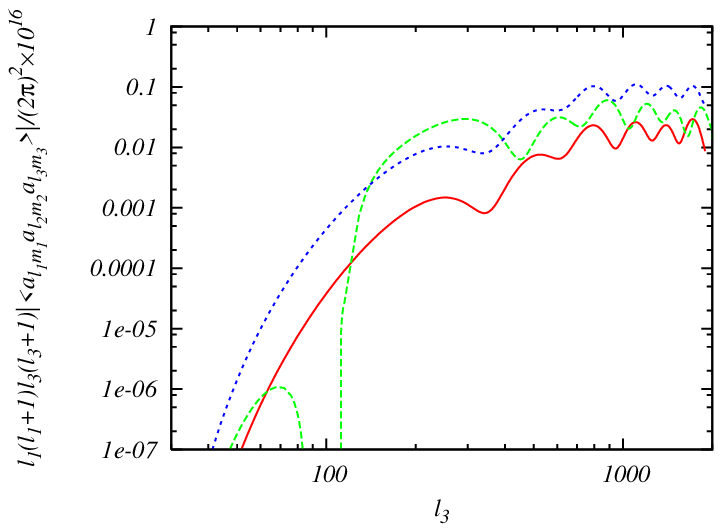}
  \end{center}
\end{minipage}
\end{tabular}
\caption{(color online) Absolute values of the CMB statistically anisotropic bispectra of the intensity mode given by Eq.~(\ref{eq:cmb_bis}) for $(m_1, m_2, m_3) = (0,0,0)$ (left figure) and $(10,20,-30)$ (right one) as the function with respect to $\ell_3$. The lines correspond to
 the spectra for $(\ell_1, \ell_2) = (102 + \ell_3, 100)$ (red solid line),
 $(|100 - \ell_3| - 2, 100)$ (green dashed line) and $(100 + \ell_3, 100)$ (blue dotted line). The parameters are identical to the values defined in
 Fig.~\ref{fig:SSS_RV_samel.eps}.}
  \label{fig:SSS_RV_C1_difl.eps}
\end{figure}

Figure \ref{fig:SSS_RV_C1_difl.eps} shows the CMB anisotropic bispectra of the intensity mode given by Eq.~(\ref{eq:cmb_bis}) with $C = 1$ for the several configurations of $\ell$'s and $m$'s as a function of $\ell_3$. The red solid line and green dashed line satisfy the special relation (\ref{eq:special_l}), namely, $\ell_1 = \ell_2 + \ell_3 + 2, |\ell_2 - \ell_3| - 2$, and the blue dotted line obeys a configuration of Eq.~(\ref{eq:default_l}), namely, $\ell_1 = \ell_2 + \ell_3$. From this figure, we confirm that the signals in the special configuration (\ref{eq:special_l}) are comparable in magnitude to those for $\ell_1 = \ell_2 + \ell_3$. Therefore, if the rotational invariance is violated on the primordial bispectrum of curvature perturbations, the signals for $\ell_1 = \ell_2 + \ell_3 + 2, |\ell_2 - \ell_3| - 2$ can also become beneficial observables. 
Here, note that the anisotropic bispectra in the other special
configurations: $\ell_1 = \ell_2 + \ell_3 + 4, |\ell_2 - \ell_3| - 4$
are zero. It is because 
these signals arise from only the contribution of $L = L' = L_A = 2, L_1 = \ell_1 \pm 2, L_2 = \ell_2 \pm 2$ in Eq.~(\ref{eq:cmb_bis}) owing to the selection rules of the Wigner symbols, and the summation of the four Wigner-$3j$ symbols over $M$ vanishes for all $\ell$'s and $m$'s. Hence, in this anisotropic bispectrum, the additional signals arise from only two configurations $\ell_1 = \ell_2 + \ell_3 + 2, |\ell_2 - \ell_3| - 2$ and these two permutations.

\section{Summary and discussion}

In this paper, we investigated the statistical anisotropy in the CMB bispectrum by considering the modified hybrid inflation model where the waterfall field also couples with the vector field \cite{Yokoyama:2008xw}.
We calculated the CMB bispectrum sourced from the non-Gaussianity of curvature perturbations affected by the vector field. 
In this inflation model, owing to the dependence on the direction of the vector field, the correlations of the curvature perturbations violate the rotational invariance. Then, interestingly, even if the magnitude of the parameter $g_\beta$ characterizing the statistical anisotropy of
the CMB power spectrum is too small, the amplitude of the non-Gaussianity can become large depending on several coupling constants of the fields. 

Following the procedure of Ref.~\citen{Shiraishi:2010kd}, we formulated the statistically anisotropic CMB bispectrum and confirm that three azimuthal quantum numbers $m_1, m_2, m_3$ are not confined only to the Wigner symbol 
$\left(
  \begin{array}{ccc}
  \ell_1 & \ell_2 & \ell_3 \\
  m_1 & m_2 & m_3
  \end{array}
 \right)$. 
This is evidence that the rotational invariance is violated in the CMB bispectrum and implies the existence of the signals not obeying the triangle condition of the above Wigner symbol as $|\ell_2 - \ell_3| \leq \ell_1 \leq \ell_2 + \ell_3$. 
We demonstrated that the signals of the CMB bispectrum for $\ell_1 = \ell_2 + \ell_3 + 2, |\ell_2 - \ell_3| -2$ and these two permutations do not vanish. In fact, the statistically isotropic bispectra are exactly zero for these configurations; hence, these signals have the pure information of the statistical anisotropy. 
Because the amplitudes of these intensity bispectra are comparable to those for $\ell_1 = \ell_2 + \ell_3$, it might be possible to detect these contributions of the statistical anisotropy in future experiments, which would give us novel information about the physics of the early Universe.
Of course, also for the $E$-mode polarization, we can give the same discussions and results. 

Although we assume a specific potential of inflation to show the statistical anisotropy on the CMB bispectrum explicitly, the above calculation and discussion will be applicable to other inflation models where the rotational invariance violates.

\section*{Acknowledgements}
We would like to thank Mindaugas Karciauskas for notifying us of a mistake and Jiro Soda for his useful comments. 
This work is supported in part by the Grant-in-Aid for JSPS Research under
 Grant No.~22-7477 (M. S.), JSPS Grant-in-Aid for Scientific Research under Grant No.~22340056 (S. Y.), Grant-in-Aid for Scientific Research on Priority Areas No. 467 ``Probing the Dark Energy through an Extremely Wide and Deep Survey with Subaru Telescope'', and Grant-in-Aid for Nagoya University Global COE Program ``Quest for Fundamental Principles in the Universe:
 from Particles to the Solar System and the Cosmos,'' from the Ministry
 of Education, Culture, Sports, Science and Technology of Japan. 
We also acknowledge the Kobayashi-Maskawa Institute for the Origin of
Particles and the Universe, Nagoya University for providing computing
resources useful in conducting the research reported in this paper.


\end{document}